\documentclass[11pt,a4paper]{article}
\usepackage{charter,url}
\usepackage[utf8]{inputenc}
\usepackage[bottom=3cm,top=3cm,left=3cm,right=3cm]{geometry}
\usepackage{graphicx}
\usepackage{amsmath}
\usepackage{mathrsfs}
\usepackage{amsfonts}
\usepackage{amssymb}
\usepackage{bbm}
\usepackage{endnotes}
\usepackage{hyperref}
\usepackage{color}

\newcommand{\be}{\begin{equation}}
\newcommand{\ee}{\end{equation}}
\newcommand{\ben}{\begin{equation*}}
\newcommand{\een}{\end{equation*}}

\newcommand{\rd}{\mathcal}

\newcommand{\badat}{\begin{alignedat}}
\newcommand{\eadat}{\end{alignedat}}
\newcommand{\bitm}{\begin{itemize}}
\newcommand{\eitm}{\end{itemize}}
\newcommand{\bmat}{\begin{pmatrix}}
\newcommand{\emat}{\end{pmatrix}}
\newcommand{\ba}{\begin{align}}
\newcommand{\bas}{\begin{align*}}
\newcommand{\ea}{\end{align}}
\newcommand{\bse}{\begin{subequations}}
\newcommand{\ese}{\end{subequations}}

\newcommand{\om}{\omega}
\newcommand{\ep}{\epsilon}
\newcommand{\virg}{\hspace{1 mm}, \hspace{8 mm}}

\begin{document}


\begin{center}

\noindent{{\LARGE{Holographic entropy of Warped-AdS$_3$ black holes}}}

\smallskip
\smallskip

\smallskip
\smallskip
\smallskip
\smallskip
\noindent{\large{Laura Donnay$^{1}$, Gaston Giribet$^{1,2,3}$}}

\smallskip
\smallskip

\end{center}

\smallskip
\smallskip
\centerline{$^1$ Universit\'{e} Libre de Bruxelles and International Solvay Institutes}
\centerline{{\it ULB-Campus Plaine CPO231, B-1050 Brussels, Belgium.}}

\smallskip
\smallskip
\centerline{$^2$ Departamento de F\'{\i}sica, Universidad de Buenos Aires and IFIBA-CONICET}
\centerline{{\it Ciudad Universitaria, Pabell\'on 1, 1428, Buenos Aires, Argentina.}}

\smallskip
\smallskip
\centerline{$^3$ Instituto de F\'{\i}sica, Pontificia Universidad Cat\'{o}lica de
Valpara\'{\i}so}
\centerline{{\it Casilla 4059, Valpara\'{\i}so, Chile.}}

\bigskip

\bigskip

\bigskip

\bigskip

We study the asymptotic symmetries of three-dimensional Warped Anti-de Sitter (WAdS) spaces in three-dimensional New Massive Gravity (NMG). For a specific choice of asymptotic boundary conditions, we find that the algebra of charges is infinite-dimensional and coincides with the semidirect sum of Virasoro algebra with non-vanishing central charge and an affine $\hat{u}(1)_k$ Ka\v{c}-Moody algebra. We show that the WAdS black hole configurations organize in terms of two commuting Virasoro algebras. We identify the Virasoro generators that expand the associated representations in the dual Warped Conformal Field Theory (WCFT) and, by applying the Warped version of the Cardy formula, we prove that the microscopic WCFT computation exactly reproduces the entropy of black holes in WAdS space.

\section{Introduction}

Warped Anti-de Sitter (WAdS) spaces are examples of the so-called non-AdS holography; that is, a proposal to generalize AdS/CFT holographic duality to cases in which the gravity side is not given by an asymptotically Anti-de Sitter spaces (AdS) space, but rather by a deformation of it. WAdS$_3$ spaces are squashed or stretched deformations of AdS$_3$ \cite{Sandinista} and have very interesting applications \cite{application1, application3, application4}. One of the most salient properties of these spaces is the fact that they admit black holes \cite{Moussa}. This permits to explore the black hole physics from the holographic point of view in a setup that goes beyond the asymptotically AdS examples.

In the recent years, different proposals for a WAdS$_3$/CFT$_2$ correspondence have been explored \cite{Aninnos, application2, DHH}. One of such proposals, dubbed WAdS/WCFT, states that asymptotically WAdS$_3$ geometries, including black holes, are dual to what has been called a {\it warped conformal field theory} (WCFT), i.e. a peculiar type of scale invariant two-dimensional theory that lacks of Lorentz invariance. In \cite{DHH}, this realization was studied in the case of Topologically Massive Gravity (TMG) and String Theory. Here, we will discuss WAdS$_3$/CFT$_2$ correspondence in a new setup, namely in the context of the parity-even three-dimensional massive gravity known as New Massive Gravity (NMG). We will give strong evidence supporting the dual description of quantum gravity about WAdS$_3$ spaces in terms of the WCFT$_2$ description.

We will study the asymptotic symmetries of WAdS$_3$ in NMG and we will find that the asymptotic symmetry algebra is infinite-dimensional and coincides with the semidirect sum of Virasoro algebra with non-vanishing central charge and an affine $\hat{u}(1)_k$ Ka\v{c}-Moody algebra. We will identify the Virasoro generators that organize the states associated to the WAdS$_3$ black hole configurations, and by applying the WCFT$_2$ version of the Cardy formula proposed in \cite{DHH}, we will prove that the microscopic WCFT$_2$ computation exactly reproduces the entropy of the WAdS$_3$ black holes. In addition, we will present a succinct derivation of such entropy formula from the CFT$_2$ point of view.

The paper is organized as follows: In Section 2, we briefly review NMG theory. In Section 3, we review the geometry of WAdS$_3$ space and the properties of asymptotically WAdS$_3$ black holes. In Section 4, we study the asymptotic isometries in WAdS$_3$ spaces and compute the algebra of charges associated to the asymptotic Killing vectors, which is found to be the semidirect sum of Virasoro algebra and the affine $\hat{u}(1)_k$ Ka\v{c}-Moody algebra. We also study the representations of this conformal algebra and identify the states that correspond to the black hole configurations in the bulk. In Section 5, we show how the black hole entropy is reproduced by the WCFT dual computation. Section 6 contains our conclusions.

\section{New Massive Gravity}

New Massive Gravity (NMG), proposed in Ref. \cite{NMG}, is a parity-even theory of gravity in three dimensions which, when linearized around maximally symmetric 
backgrounds, coincides with massive spin-2 Fierz-Pauli action. Therefore, at a generic point of the space of parameters, it propagates two local degrees of freedom. 

NMG is defined by the action
\begin{equation}
{\mathcal I} = \frac{1}{16\pi G} \int d^3 x \sqrt{-g} \left( R -2\Lambda + \frac{1}{m^2} (R_{\mu \nu }R_{\mu \nu }-\frac{3}{8} R^2)\right), \label{I}
\end{equation}
where $m$ represents the mass of the graviton. The relative coefficient $3/8$ between the two quadratic terms is essential for the theory to be free of ghosts about physically sensible backgrounds.

The equations of motion derived from (\ref{I}) take the form
\begin{equation}
R_{\mu\nu}-\frac{1}{2}R g_{\mu\nu} +\Lambda g_{\mu\nu} +\frac{1}{m^2}K_{\mu\nu} = 0 , \label{eom}
\end{equation}
where tensor $K_{\mu\nu}$ contains four derivatives of the metric (see \cite{NMG} for an explicit expression).  

NMG equations of motion (\ref{eom}) admit a large set of interesting solutions, including Schr\"odinger spaces, Lifshitz spaces, Warped AdS$_3$, and AdS$_2 \times \mathbb{R}$ spaces. Therefore, this is a fruitful arena to explore different proposals of non-AdS holography. Among them, here we will be concerned with the so-called WAdS$_3$ spaces.

\section{Warped AdS$_3$ Spaces}

As said, WAdS$_3$ spaces are solutions of NMG \cite{Clement}. These geometries are squashed or stretched deformations of AdS$_3$ space \cite{Sandinista}. We review 
these geometries below. 

\subsection{Timelike WAdS$_3$ space}

To organize the discussion in a convenient way, let us begin by considering the timelike WAdS$_3$ space. This geometry is important for our discussion as it will be ultimately associated to 
the vacuum of the WAdS$_3$ black hole spectrum we are interested in. 

The metric of timelike WAdS$_3$ corresponds to the three-dimensional G\"odel spacetime \cite{Glenn}. Its metric is given by
\begin{equation}
	d{s}^2 =-dt^2  -4\om r dt d\phi +  \frac{\ell^2 dr^2}{\left( {2r^2} (\om^2 \ell^2 +1) +2\ell^2 r \right)}- \left( \frac{2r^2}{\ell^2} (\om^2 \ell^2 -1) 
-2 r \right)d\phi^2 ,
\label{godel}
\end{equation}
and solves NMG equations of motion for the particular choice of parameters 
\begin{equation}
m^2 =-\frac{(19 \om^2 \ell^2-2)}{2\ell^2}, \ \ \ \ \ \ \  \Lambda=-\frac{(11 \om^4 \ell^4 + 28 \om^2 \ell^2 - 4)}{2\ell^2(19 \om^2 \ell^2 - 2)  }. \label{couplings}
\end{equation}

The mass of G\"odel spacetime in NMG has been recently computed in \cite{nos}, and the result was found to be
\begin{equation}
\rd M_{\text{G\"od}} = - \frac{4\ell^2\omega^2}{G(19\ell^2\omega^2-2)} \label{masagodel}.
\end{equation}

The isometry group of WAdS$_3$ space (\ref{godel}) is $SL(2,\mathbb{R})\times U(1)$, which is
generated by the four Killing vectors that the three-dimensional section of four-dimensional G\"{o}del solution
exhibits. In the particular case $\omega ^2 \ell^2=1$, solution (\ref{godel}) coincides with AdS$_3$. 

\subsection{WAdS$_3$ Black Holes}

Now, let us move to the analysis of the spacelike WAdS$_3$ spaces. In particular, we will be interested in the black hole geometries found in \cite{Moussa, Clement}, which at large distance asymptote squashed spacelike WAdS$_3$ space. The metric of these black holes is
\be
\badat{2}
\frac{ds^2}{l^2}=  dt^2 + \frac{dr^2}{(\nu^2+3)(r-r_+)(r-r_-)}+(2\nu r -\sqrt{r_+r_-(\nu^2+3)})dt d\varphi\\
+\frac{r }{4}\left[3(\nu^2-1)r+(\nu^2+3)(r_++r_-)-4\nu \sqrt{r_+r_-(\nu^2+3)}\right]d\varphi^2,
\label{warped}
\eadat
\ee
and solves NMG equations of motion for the values of parameters
\begin{equation}
m^2 = -\frac{(20\nu^2-3)}{2l^2}, \ \ \ \ \Lambda = -\frac{m^2(4\nu^4-48\nu^2+9)}{(400\nu^4-120\nu^2+9)}.
\end{equation}

The conserved charges of WAdS$_3$ black holes have been computed by different methods \cite{Clement, Goya, NPY}. The mass is given by
\be 
\rd M =Q_{\partial_t}=\frac{\nu (\nu^2+3) }{ G l(20 \nu^2 -3)} \left((r_-+r_+)\nu - \sqrt{r_+r_-(\nu^2+3)}\right) , \label{M}
\ee
while the angular momentum is given by
\be
\rd J =Q_{\partial_{\varphi }}=\frac{\nu (\nu^2+3)}{4G l(20 \nu^2 - 3)} \left( (5\nu^2 +3) r_+ r_- - 2 \nu \sqrt{r_+r_-(\nu^2+3)}(r_+ + r_-)\right). \label{J}
\ee

Black holes (\ref{warped}) include extremal configurations, corresponding to $r_+=r_-$. In those cases, the angular momentum saturates the condition
\begin{equation}
\rd J  \leq \frac{Gl(20\nu^2-3)}{4\nu(\nu^2+3)} \rd M ^2 , \label{ext}
\end{equation}
which is the necessary condition for the existence of horizons. Condition (\ref{ext}) is supplemented with $\rd M \geq 0$.

Black hole solutions (\ref{warped}) are obtained from the timelike WAdS$_3$ space (\ref{godel}) by means of global identifications \cite{Aninnos}, in the same way as BTZ black holes \cite{BTZ} are obtained from AdS$_3$ as 
discrete quotients \cite{BHTZ}.

This orbifold construction preserves a $U(1) \times U(1)$ subgroup of $SL(2,\mathbb{R})\times U(1)$ isometries, which is generated by the two Killing vectors
\begin{equation}
	\xi^{(1)} =  \partial_t \ , \ \ \ \ \xi^{(2)} = \frac{2l\nu }{(\nu^2+3)} \partial_t + \partial_{\varphi} \, .
\label{RemainingKV}
\end{equation}

The global identifications, generated by to Killing vectors (\ref{RemainingKV}), generate two periods $\beta_R$ and $\beta_L$. The inverse of these periods yield the 
two geometrical temperatures
\begin{eqnarray}
T_R &=& \beta_R^{-1} = \frac{(\nu^2+3)}{8\pi l^2} (r_+ - r_-) , \\
T_L &=& \beta_L^{-1} = \frac{(\nu^2+3)}{8\pi l^2} (r_+ + r_- -\frac{1}{\nu } \sqrt{(\nu^2+3)r_- r_+}) .
\end{eqnarray}

The entropy is given by 
\begin{equation}
S_{\text{BH}} = \frac{8\pi \nu^3}{(20\nu^2 - 3)G} (r_+ - \frac{1}{2\nu }\sqrt{(\nu^2+3)r_-r_+}) , \label{S}
\end{equation}

and reads, in terms of the charges (\ref{M})-(\ref{J}),
\begin{equation}
S_{\text{BH}} = \frac{4\pi l\nu }{(\nu^2 + 3)} (\rd M +\sqrt{\rd M^2 -k\rd J}), \label{V2}
\end{equation}
where $ k =  {4 \nu (3+\nu^2)}/{(G l(20 \nu^2 -3))} $. This way of writing the entropy will be important for our purpose.

\section{Asymptotic symmetries}
\subsection{Asymptotic isometry algebra}
In this section, we will study the notion of asymptotically WAdS$_3$ spaces. To do this, first we choose as a background metric, $ g$, the solution \eqref{warped} with $r_+=0=r_-$; and then we impose the same boundary conditions as in \cite{Compere3}, namely\footnote{In WAdS$_3$ spaces, other sets of boundary conditions have been considered; see for instance \cite{Enoc, SongCompere}. It would be interesting to investigate other definitions of boundary conditions in the context of NMG as well.}
\be
\badat{2}
&g_{tt}=l^2+\rd O(r^{-1}) \virg g_{tr}=\rd O(r^{-2}) \virg g_{t\varphi}=l^2 \nu r+\rd O(1), \\
&g_{rr}=\frac{l^2}{(\nu^2+3)r^2}+\rd O(r^{-3}) \virg g_{r\varphi}=\rd O(r^{-1}) \virg g_{\varphi \varphi}=\frac{3}{4}r^2 l^2(\nu^2-1)+\rd O(r),
\label{falloff}
\eadat
\ee
which include in particular the black hole solutions \eqref{warped}. The set of asymptotic diffeomorphisms allowed by these boundary conditions are
\be
\badat{2}
&\ell_n= (N_1 \ e^{in\varphi} +\rd O(r^{-1})) \partial_t+(-in r  e^{in\varphi}+\rd O(1)) \partial_r+ ( e^{in\varphi}+\rd O(r^{-2})) \partial_\varphi,  \\
&t_n =(N_2 \ e^{in\varphi}+\rd O(r^{-1})) \partial_t,  
\label{asdiff}
\eadat
\ee
where $n \in \mathbb Z$, and where $N_1, N_2$ are two arbitrary normalization constants. Indeed, acting with $\ell_n, t_n$ on a metric obeying \eqref{falloff} leads to a perturbation obeying the same falling-off conditions. 

The generators \eqref{asdiff} satisfy the algebra
\be
\badat{3}
&i[\ell_m,\ell_n]=(m-n)\ell_{m+n}, \\
&i[\ell_m,t_n]=-n t_{m+n}, \\
&i[t_m,t_n]=0. 
\eadat
\ee

This is the semidirect sum of Witt algebra and the loop algebra of $u(1)$.

\subsection{Algebra of charges}

In the covariant formalism \cite{Barnich:2001jy,Barnich:2007bf}, conserved charges associated to an asymptotic Killing vector $\xi$ are given, in three spacetime dimensions, by the expression 
\be
\delta Q_\xi [\delta g, g]=\frac{1}{16 \pi G} \int_{0}^{2\pi} \sqrt{-g}\, \ep_{\mu \nu \varphi} \,k_\xi^{\mu \nu}[\delta g, g]d\varphi,
\label{formulacharge}
\ee
with $g$ a solution, $\delta g$ a linearized metric perturbation around it, and $k_\xi^{\mu \nu}[\delta g, g]$ a one-form potential of the linearized theory.

This potential depends of the theory considered and was computed for NMG in Ref. \cite{NPY} for Killing vectors $\xi$ using the so-called Abbott-Deser-Tekin formalism. The result takes the form
\be
k_{\xi}^{\mu \nu}=Q^{\mu \nu}_R+\frac{1}{2m^2}Q^{\mu \nu}_{K},
\label{kexact}
\ee
where the first contribution comes from the pure GR part of the equations of motion, while $Q^{\mu \nu}_{K}=Q^{\mu \nu}_{R_2}-\frac{3}{8}Q^{\mu \nu}_{R^2}$ accounts for the contribution of the $K_{\mu \nu}$ tensor. Explicit expressions for $Q^{\mu \nu}_{R}$, $Q^{\mu \nu}_{R^2}$, and $Q^{\mu \nu}_{R_2}$ can be found in equations (13), (22) and (28) in \cite{NPY}, respectively. As an example, the black hole mass and angular momentum (\ref{M})-(\ref{J}) are obtained by computing charges associated to the Killing vectors $\partial_t$ and $\partial_{\varphi }$ respectively.

In the case of asymptotic Killing vectors, the one-form potential to be considered is given by $k_{\xi}[\delta g, g]+k^S_{\xi}[\delta g, \rd L_{\xi} g]$, where the second term is a supplementary contribution linear in the Killing equation and its derivatives. This term is at the origin of the difference between the conserved charges in the Barnich-Brandt-Comp\`ere formalism \cite{Barnich:2001jy,Barnich:2007bf} and in covariant phase space methods \cite{IyerWald}. However, in most of the cases, this term does not contribute to any charge. For instance, in the case of the WAdS black hole solution (\ref{warped}) and with the asymptotic Killing vectors \eqref{asdiff}, this term has been shown to be of order $\rd O(r^{-1})$ in Topologically Massive Gravity (TMG) \cite{Compere2}. A way to see that it will not contribute to the asymptotic charge neither here is to notice that the piece coming from the $K_{\mu \nu }$ tensor of NMG can only contain terms proportional to the second derivative of the Killing equation, and therefore will be at most of order $\rd O(r^{-1})$, as it happens in TMG. The consistency of the results confirms this a posteriori.

If we denote the charges differences between the black hole solution \eqref{warped} and the background $ g$ by $L_n=Q_{\ell_n}$, $T_n=Q_{t_n}$, we find the following charge algebra
\be
\badat{3}
&i\{L_m,L_n\}=(m-n)L_{m+n}+\frac{1}{12}(c m^3 + 6 k  N^2_1 m)\delta_{m+n,0} ,  \\
&i\{L_m,T_n\}=-n T_{m+n}+\frac{k}{2} N_1 N_2 m\delta_{m+n,0},  \\
&i\{T_m,T_n\}=  \frac{k}{2} N_2^2 m \delta_{m+n,0}, 
\label{74}
\eadat
\ee
where 
\be
c=-\frac{96 l \nu^3}{G (20 \nu^4 +57 \nu^2  -9)} 
\label{c}
\ee
and
\be
k =  \frac{4 \nu (3+\nu^2)}{G l(20 \nu^2 -3)} .
\label{k}
\ee

Algebra (\ref{74}) is equivalent to the semidirect sum of Virasoro algebra with central charge $c$ and the affine $\hat{u}(1)_k$ Ka\v{c}-Moody algebra of level $k$. 


Normalization $N_1$ in (\ref{74}) can be fixed by matching the asymptotic Killing vector $\ell_0$ with the vector $\xi^{(2)}$ in (\ref{RemainingKV}), which yields
\begin{equation}
N_1=\frac{2l\nu }{(\nu^2+3)}. \label{tornado1}
\end{equation}
Analogously, for $t_0$ to match $\xi^{(1)}$ in (\ref{RemainingKV}), we fix $N_2=1$ .

Defining the new generators
\begin{equation}
P_n \equiv T_n + \frac{k }{2} N_1 N_2 \ , \label{tornado4}
\end{equation}
and realizing that $c=-6kN_1^2 $, algebra (\ref{74}) takes the familiar form
\be
\badat{3}
&i\{L_m,L_n\}=(m-n)L_{m+n}+\frac{c}{12}(m^3 - m)\delta_{m+n,0} ,  \\
&i\{L_m,P_n\}=-n  P_{m+n} ,  \\
&i\{P_m,P_n\}= \frac{k}{2} m\delta_{m+n,0}. 
\label{tornado6}
\eadat
\ee

Note that, the absolute value of (\ref{c}) coincides with the value of the central charge conjectured in \cite{Goya}, which leads to reproduce the entropy of WAdS$_3$ black holes (\ref{warped}); namely
\begin{equation}
S= \frac{\pi^2 l}{3} c (T_R + T_L) .
\end{equation}
If $\nu /G >0$, then we find 
\begin{equation}
c<0 \ \ \ \text{and} \ \ \  k>0.  \label{conditionsck}
\end{equation}
In the next section, we will discuss (\ref{conditionsck}) in the context of the unitary highest-weight representations of the algebra (\ref{tornado6}). However, let us mention here that, since $L_0$ is not bounded from below, one can freely reverse the sign of $c$ by redefining the generators as $L_n \to -L_{-n}$.

Before concluding this section, let us mention that the computation of asymptotic charges can also be carried out in the case of timelike WAdS$_3$ spaces. The algebra obtained in that case is the same, with the central charge $c$ and the level $k$ given by 
\begin{equation}
c=\frac{48 \ell^4 \om^3}{G (19 \ell^4 \om^4 +17 \ell^2 \om^2 -2)}  \ , \ \ \ \ \ 
k =  \frac{8\om (1+\ell^2 \om^2)}{G (19 \ell^2 \om^2 -2)}. 
\end{equation}
respectively, where $\omega = \nu/l$ and $\omega^2\ell^2+2=3\ell^2/l^2$. As a consistency check of this result, one can observe that the value of $c$ tends to the AdS$_3$ value $c_{\text{AdS}} = 3\ell/(2G)(1+1/(2m^2\ell^2))$ in the limit $\omega^2\ell^2=1$.

\subsection{Unitary highest-weight representations}

Algebra (\ref{74}) admits a simple automorphism, given by the spectral flow transformation
\begin{equation}
P_n \to \tilde{P}_n = P_n + p_{0} \delta_{n,0} , \label{spectralflow}
\end{equation}
with $p_{0} $ being an arbitrary complex number. This one-parameter transformation, which in the case of $\hat{u}(1)_k$ algebra merely amounts to shift the 
zero-mode of $P_n$, has to be taken into account when building up the highest-weight representations. 

We can now play the standard game and promote charges $L_n$ and $P_n$ to the rank of operators acting on a vector space whose elements are represented by quantum 
states $|v\rangle $. This amounts to replace the Poisson brackets in (\ref{tornado6}) by commutators, namely $i\{ ,\} \to [,]$. In addition, for these operators 
we have the hermiticity relations
\begin{equation}
P_n^{\dagger }=P_{-n} , \ \ \ \ \ \ L_n^{\dagger }=L_{-n}.
\end{equation}
Our informal style prevents us from using hats.

Since, in particular, $[L_0,\tilde{P}_0] =0$, then one can construct the highest-weight representations starting with the primary states $|v\rangle = 
| h,p,p_{0} \rangle $, labeled by three complex parameters $h$, $p$, $p_{0} $ corresponding to the eigenvalues
\begin{eqnarray}
L_0 | h,p,p_{0} \rangle = h \ | h,p,p_{0} \rangle  ,  \ \ \ \ \ 
\tilde{P}_0 | h,p,p_{0} \rangle = p \ | h,p,p_{0} \rangle , 
\end{eqnarray}
and imposing
\begin{eqnarray}
L_n | h,p,p_{0} \rangle = 0 , \ \ \ \ \ \ \ 
\tilde{P}_n | h,p,p_{0} \rangle = 0 , \ \ \ \ \ \forall n > 0
\end{eqnarray}
where $p_{0} $ refers to which spectrally flowed sector the state corresponds to. For instance, the state of the $p_{0} =0$ sector obeys $P_0 | h,p,0 \rangle = 
p | h,p,0 \rangle$ and $\tilde{P}_0 | h,p,0 \rangle = (p+p_{0} ) | h,p,0 \rangle$, where $\tilde{P}_0$ is defined as in (\ref{spectralflow}). This invites to 
identify states $| h,p-p_{0} ,0 \rangle $ with states $| h,p ,p_{0} \rangle $ for all $p_{0} $. This seems trivial in the case of $\hat{u}(1)_k$ affine algebra, 
but spectral flow acts in a non-trivial way on algebras such as $\hat{su}(2)_k$ or $\hat{sl}(2)_k$, of which $\hat{u}(1)_k$ is a subalgebra, mapping in the 
former cases Ka\v{c}-Moody primary states to descendents and, in the case $\hat{sl}(2)_k$, generating new representations.

Descendent states are then defined by acting on primaries $|h,p,p_{0} \rangle $ with arrays of positive modes $P_{-n}$ and $L_{-n}$ with $n\geq 0$. 

Unitarity constraints are derived from algebra (\ref{tornado6}) in the usual way. In particular, this yields the conditions on $c$ and $k$, together with 
the dimension $h$ and momentum $p_{0} $ of the states. More precisely, demanding $\| L_{-1}| h,p,p_{0} \rangle \|^2 \geq 0$ yields $h\geq 0$; analogously, $\| 
P_{0}| h,p,p_{0} \rangle \|^2 \geq 0$ implies $p\in \mathbb{R}$. Spectral flow symmetry (\ref{spectralflow}) also implies $p_{0}\in \mathbb{R}$. On the other 
hand, positivity of $\| L_{-n}| h,p,p_{0} \rangle \|^2 $ (for large $n$) and $\| P_{-1}| h,p,p_{0} \rangle \|^2 $ yields
\begin{equation}
c>0 \ \ \ \text{and} \ \ \  k\geq 0,  \label{conditionsck2}
\end{equation}
respectively. These conditions seem to be in contradiction with (\ref{conditionsck}). A priori, this may seen puzzling; however, this is not a problem at this point 
because $L_0$ is not bounded from below\footnote{Here, it is probably convenient to be reminded that Virasoro algebra is invariant under the inversion $L_{n} \to -L_{-n}$ and $c\to -c$; see (\ref{posta}).}. Instead, the Virasoro operators associated to the black hole spectrum ((\ref{TY1}) and (\ref{posta}) below) will be bounded from below and have positive central charges (see (\ref{TY2}) below). To see this, let us first define
\begin{equation}
L^{-}_n \equiv  \frac{1}{k} \sum_{m} : P_{-n-m} P_{m} : , \label{TY1}
\end{equation}
where $: \ :$ stands for {\it normal ordering}. Operators $L^{-}_n$ obey Virasoro algebra\footnote{Notice also that if one applies spectral flow transformation 
(\ref{spectralflow}), one verifies that the zero mode changes as follows $
L^{-}_0 \to L^{-}_0 - \frac{2p_{0} }{k} P_0^2 - \frac{p_{0}^2}{k} .$} with $c_-=1$; namely
\begin{equation}
[ L^{-}_m , L^{-}_n ] = (m-n) L^{-}_{m+n} + \frac{1}{12} m( m^2-1) \delta_{m+n,0},
\end{equation}
and satisfy
\begin{equation}
[L^{-}_m,P_n]=-n  P_{-m+n} . \label{beb}
\end{equation}
This is nothing but the Sugawara construction in the case of $\hat{u}(1)_k$; see also \cite{blago}. 

Secondly, we define operators
\begin{equation}
L_{n}^{+} \equiv L^{-}_{n} - L_{-n} , \label{posta}
\end{equation}
which also generate a Virasoro algebra.

Notice from (\ref{tornado6}) and (\ref{beb}) that operators $L^{+}_n$ commute with $P_m$ and, consequently, one finds two commuting Virasoro algebras; namely
\begin{equation}
[L^{+}_n,L^{-}_m] = 0.
\end{equation}

In addition, unlike what happens with Virasoro algebra generated by $L_n$, operators $L_n^{\pm }$ evaluated on the black hole spectrum turn out to be bounded 
from below. To see this explicitly, notice that the energy spectrum $L^{\pm }_0$ is given by
\begin{equation}
h^{+} = \frac{1}{k} \rd M ^2 -\rd J - \frac{c}{24} , \ \ \ \ h^{-} = \frac{1}{k} (\rd M +kN_1 /2)^2,
\end{equation}
where $h^{\pm }$ refer to the eigenvalue of $L_0^{\pm}$, and where we have used that $c=-6kN_1^2$. On the other hand, from (\ref{ext}) we observe that the black hole spectrum is such that both $L_0^{+}$ and $L_0^{-}$ are bounded from below. In fact, from (\ref{ext}) we have 
\begin{equation}
\frac{\rd M^2}{k} - \rd J = \frac{\nu^2 k}{16} (r_+ - r_-)^2 \geq 0 , \label{bound}
\end{equation}
which implies the bounds
\begin{equation}
L^{\pm }_0 \geq  - \frac{c}{24} . \label{TY2}
\end{equation}

In the next subsection we will see how the microstates representing black hole configurations (\ref{warped}) seem to organize themselves in representations of Virasoro algebras generated by $L_n^{\pm }$. Strong evidence of that is the CFT$_2$ rederivation of the black hole entropy (\ref{S}).

\section{WCFT$_2$ and microscopic entropy}

In Ref. \cite{DHH}, a Cardy type formula for WCFT has been proposed. This formula is supposed to give the asymptotic growth of states in the dual theory, which 
would lack of full Lorentz invariance. We will describe below how such a formula actually comes from the two Virasoro algebras generated by $L^{\pm }_n$. 

Following 
\cite{DHH}, we first shift the Virasoro operator $L^{+ }_0$ as follows
\begin{equation}
L_0^+ \to \tilde{L}_{0}^+ = L_{0}^+ + \frac{c}{24}, \label{A}
\end{equation}
and perform the spectral flow operation
\begin{equation}
P_0 \to \tilde{P}_0 = P_0 -\frac{kN_1}{2}. \label{B}
\end{equation}
Notice that performing (\ref{A}) and (\ref{B}) corresponds to 
having chosen in (\ref{asdiff}) the gauge $N_1=0$; this is exactly what is done in \cite{DHH} (see (77) therein, cf. (14), (19) and (20) in \cite{Compere}). Physically, it corresponds to associate the zero-energy state $\tilde{L}_0^{\pm }=0$ to the zero-mass black hole $\rd M = \rd J = 0$.

Now, we are ready to show how the Cardy formula associated to the new Virasoro operators (\ref{A}) and (\ref{B}) actually reproduces the black hole entropy. First, we write the standard CFT Cardy formula 
\begin{equation}
S_{\text{CFT}} = 2\pi \sqrt{-4\tilde{L}_0^{- (\text{vac})} \tilde{L}_0^{-} } + 2\pi \sqrt{-4\tilde{L}_0^{+ (\text{vac})} \tilde{L}_0^{+} }, \label{jk}
\end{equation}
where $\tilde{L}_0^{\pm (\text{vac})}$ correspond to the minimum values of $\tilde{L}_0^{\pm}$, i.e. the value of the {\it vacuum geometry}. It is important to remark that the way of writing Cardy formula in (\ref{jk}) admits the possibility of the spectrum of $\tilde{L}_0^{\pm }$ to exhibit a gap with respect to the value $-c^{\pm }/24$. More precisely, it takes into account that in theories with such a gap, the saddle point approximation involved in the derivation of the Cardy formula yields an effective central charge $c_{\text{eff}}$ given by $c_{\text{eff}}/6 = -4L_0^{\text{(vac)}}$.

Recalling that $\tilde{L}_0^{-} = \tilde{P}_0^2/k$, formula (\ref{jk}) reads
\begin{equation}
S_{\text{WCFT}} = \frac{4\pi i}{k} \tilde{P}_0^{(\text{vac})} \tilde{P}_0  + 4\pi \sqrt{-\tilde{L}_0^{+(\text{vac})} \tilde{L}_0^{+} } . \label{Cardy}
\end{equation}

In turn, the only remaining ingredient needed to apply this formula is to find out which is the right {\it vacuum geometry}. Because the theory is parity even, we naturally expect the vacuum 
geometry to be\footnote{This is different from the case of TMG.} $\rd J ^{\text{(vac)}}=0$; that is to say,
\begin{equation}
\tilde{P}_0^{\text{(vac)}} =\rd M^{\text{(vac)}}  , \ \ \ \ \tilde{L}_0^{+\text{(vac)}} =  \frac{1}{k} (\rd M^{\text{(vac)}} )^2 .
\end{equation}
This yields
\begin{equation}
S_{\text{CFT}} = \frac{4\pi i}{k} \rd M^{\text{(vac)}} (\rd M +\sqrt{\rd M^2 -k\rd J}). \label{V1}
\end{equation}
And, indeed, we verify that entropy (\ref{S}) exactly matches formula (\ref{Cardy}) if one identifies the vacuum geometry with the G\"odel geometry 
(\ref{godel}); namely 
\begin{equation}
\rd M^{\text{(vac)}}  = i \rd M_{\text{G\"od}} = - i \frac{4\ell^2\omega^2}{G(19\ell^2\omega^2-2)} . \label{cp}
\end{equation}

The identification of timelike WAdS$_3$ (\ref{cp}) spacetime as the vacuum geometry of the spacelike WAdS$_3$ black hole spectrum is something that had been 
observed in \cite{DHH} for the case of TMG and String Theory. Here we obtain the similar result for the case of NMG. It is natural, on the other hand, that the vacuum geometry preserves the full $SL(2,\mathbb{R})\times U(1)$.

Therefore, we have shown that
\begin{equation}
S_{\text{BH}} = S_{\text{CFT}} . \label{matching}
\end{equation}

\section{Conclusions}

In this paper, we have computed the asymptotic symmetry algebra corresponding to Warped Anti-de Sitter (WAdS) spaces of three-dimensional New Massive Gravity (NMG). We have shown that this is given by the semi-direct sum of one Virasoro algebra (with non-vanishing central charge) and one affine $\hat{u}(1)_k$ Ka\v{c}-Moody algebra. We have identified the precise Virasoro generators that organize the states associated to the WAdS$_3$ black hole configurations, which led us to rederive the WCFT entropy formula (\ref{Cardy}) in a very succint way, starting from the standard CFT Cardy formula (\ref{jk}).

By applying the WCFT entropy formula \cite{DHH}, we have proved that the microscopic computation in the dual WCFT exactly reproduces the entropy of the WAdS$_3$ black holes. Essential ingredients for the matching (\ref{matching}) to hold are: {\it a)} The definition of Virasoro algebras generated by $\tilde{L}_{n}^{\pm}$ as in (\ref{TY1}) and (\ref{posta}) with (\ref{A}) and (\ref{B}), {\it b)} the choice $N_1=0$ in (\ref{asdiff}) that induces the shifting (\ref{A})-(\ref{B}), {\it c)} the identification of the vacuum geometry of the black hole spectrum as in (\ref{cp}). These ingredients agree with the recipe proposed in \cite{DHH} for the cases of Topologically Massive Gravity and String Theory. 

As further directions, we can mention some open questions that deserve to be studied further: {\it i)} The first one is whether new conformal structures appear for different sets of asymptotic boundary conditions. It is well known that the properties of the boundary theory depend on the different asymptotic behavior of the fields at infinity. Therefore, the question arises as to whether asymptotically WAdS$_3$ boundary conditions exist such that, while being sufficiently restrictive for the dual theory not to have negative-norm excitations, they still allow for a rich space of solutions in the bulk. Different proposals for WAdS$_3$ asymptotics were proposed in the literature \cite{Enoc, SongCompere} and it would be interesting to investigate their dynamical implications. {\it ii)} Another interesting future direction is to study the applications of WAdS$_3$/WCFT$_2$ to Kerr$_4$/CFT$_2$ correspondence. In Kerr/CFT in four (and higher) dimensions, the so-called Near Horizon Extremal Kerr (NHEK) geometry is closely related to the WAdS$_3$ spaces studied here \cite{Sandinista, Aninnos}, and, therefore, the application of the holographic ideas developed for WAdS$_3$/WCFT$_2$ to the more realistic case of four-dimensional spinning black holes is of principal interest. 


\begin{equation*}
\end{equation*}

The authors thank Sophie de Buyl, Geoffrey Comp\`{e}re, St\'ephane Detournay, and P.-H. Lambert. L.D.~is a research fellow of the ``Fonds pour la Formation \`a
la Recherche dans l'Industrie et dans l'Agriculture''-FRIA
Belgium and her work is supported in part by IISN-Belgium and by
``Communaut\'e fran\c caise de Belgique - Actions de Recherche
Concert\'ee. The work of G.G. has been partially funded by FNRS-Belgium (convention FRFC PDR T.1025.14 and convention IISN 4.4503.15), by the Communaut\'{e} Fran\c{c}aise de Belgique
through the ARC program and by a donation from the Solvay family. The support of CONICET, FNRS+MINCyT and UBA through grants PIP 0595/13, BE 13/03 and UBACyT 20020120100154BA, respectively, is greatly acknowledged.

\providecommand{\href}[2]{#2}\begingroup\raggedright\endgroup

\end{document}